\documentclass[conference]{IEEEtran}
\IEEEoverridecommandlockouts

\usepackage{cite}
\usepackage{amsmath,amssymb,amsfonts}
\usepackage{algorithmic}
\usepackage{graphicx}
\usepackage{textcomp}
\usepackage{xcolor}
\usepackage{makecell}
\usepackage{url}

\def\BibTeX{{\rm B\kern-.05em{\sc i\kern-.025em b}\kern-.08em
T\kern-.1667em\lower.7ex\hbox{E}\kern-.125emX}}

\begin{document}

\title{\LARGE A Multilingual Framework for Dysarthria: Detection, Severity Classification, Speech-to-Text, and Clean Speech Generation}

\author{
\IEEEauthorblockN{
Ananya Raghu$^{1,\ast}$,
Anisha Raghu$^{1,\ast}$,
Nithika Vivek$^{1,\ast}$,
Sofie Budman$^{1,\ast}$,
Omar Mansour$^{1,2}$
}
\IEEEauthorblockA{
\textit{$^{1}$Beaver Works Summer Institute, Massachusetts Institute of Technology} \\
\textit{$^{2}$Columbia University} \\
Emails: \texttt{ananyaraghu10@gmail.com, anisharaghu10@gmail.com}\\
\texttt{nithika.vivek@gmail.com, sofiebudman@gmail.com, omar.mansour@columbia.edu}
}
\thanks{$^{\ast}$ Equal contribution.}
}

\maketitle

\begin{abstract}
Dysarthria is a motor speech disorder that results in slow and often incomprehensible speech. Speech intelligibility significantly impacts communication, leading to barriers in social interactions. Dysarthria is often a characteristic of neurological diseases including Parkinson’s and ALS, yet current tools lack generalizability across languages and levels of severity. In this study, we present a unified AI-based multilingual framework that addresses six key components: (1) binary dysarthria detection, (2) severity classification, (3) clean speech generation, (4) speech-to-text conversion, (5) emotion detection, and (6) voice cloning. We analyze datasets in English, Russian, and German, using spectrogram-based visualizations and acoustic feature extraction to inform model training. Our binary detection model achieved 97\% accuracy across all three languages, demonstrating strong generalization across languages. The severity classification model also reached 97\% test accuracy, with interpretable results showing model attention focused on lower harmonics. Our translation pipeline, trained on paired Russian dysarthric and clean speech, reconstructed intelligible outputs with low training (0.03) and test (0.06) L1 losses. Given the limited availability of English dysarthric-clean pairs, we finetuned the Russian model on English data and achieved improved losses of 0.02 (train) and 0.03 (test), highlighting the promise of cross-lingual transfer learning for low-resource settings. Our speech-to-text pipeline achieved a Word Error Rate of 0.1367 after three epochs, indicating accurate transcription on dysarthric speech and enabling downstream emotion recognition and voice cloning from transcribed speech. Overall, the results and products of this study can be used to diagnose dysarthria and improve communication and understanding for patients across different languages.

\end{abstract}


\section{Introduction}
\subsection{Problem Statement}
Dysarthria is a motor speech disorder and a common symptom of neurological conditions such as ALS, Parkinson’s disease, stroke, and cerebral palsy \cite{mayoclinic_dysarthria_2024}. It arises when the nervous system damage impairs the muscles involved in speaking, leading to slurred, slow speech that is difficult to understand \cite{Jayaraman2025-vp}. While not a disease itself, dysarthria significantly impairs communication and quality of life, frequently leading to social isolation, misdiagnosis, or reduced access to care. Studies report that dysarthria occurs in up to 60\% of stroke patients and affects as many as 90\% of individuals with Parkinson’s disease \cite{asha_dysarthria_adults}. Despite its prevalence, Dysarthria is often under-recognized, particularly in its milder forms or in multilingual populations \cite{miller2014crosslanguage}. A recent study demonstrated that a listener's native language significantly influences their perceptual ratings of dysarthria, particularly for articulatory and rhythmic characteristics \cite{Kim2024-jj}. This highlights a fundamental limitation in human-based assessment, as a clinician's ability to accurately perceive and rate a speaker's dysarthria can be compromised when they are not a native speaker of the language. This suggests that current diagnostic methods that rely on subjective evaluations by speech-language pathologists are constrained by language familiarity and clinical access. Furthermore, they are prone to human error and bias, which can delay proper treatment, especially for early stage dysarthria \cite{HASSAN2025108128, Wolfrum2023-vk}.

In contrast, machine learning models trained on diverse, labeled datasets offer an objective alternative to human assessments. By extracting language-agnostic acoustic features and learning features across speech samples, ML can reduce diagnostic bias and enable more consistent screening. This makes machine learning based tools especially promising for accessible dysarthria detection across diverse healthcare settings. 

\subsection{Prior Machine Learning Approaches}

\subsubsection{Prior work: Dysarthria Detection}
Recent advances in machine learning have led to the development of models capable of detecting dysarthria using acoustic features such as Mel-Frequency Cepstral Coefficients (MFCCs), spectrograms, or prosodic cues. Prior work has focused largely on binary classification, distinguishing dysarthric from healthy speech, using convolutional or recurrent neural networks trained on datasets like TORGO and UA-Speech \cite{MAHENDRAN2023100913, Shih2022-ta}. However, these models are typically trained and evaluated on a single language, limiting their clinical applicability across multilingual populations. 

\subsubsection{Prior Work: Severity Classification}
While recent approaches \cite{sajiha2024automatic} to dysarthria severity classification have received high performance using neural networks, they are often black boxes, not explaining the reasoning behind model classification. Furthermore, features extracted for the model, including embeddings from wav2vec2 \cite{sajiha2024automatic}, do not offer insight on which acoustic characteristics of the slurred speech distinguish it from clean or less severe dysarthric speech. An interpretable model can thereby give insight as to which features of the speech are altered in dysarthria, helping with speech therapy and increasing patient understanding. 
\subsubsection{Prior work: Speech Synthesis}
Prior research has explored several generative voice conversion and augmentation approaches for translating dysarthric speech to regular speech. The CycleGAN-VC model architecture was applied to Korean dysarthric speech (18700) utterances and healthy controls reducing Word-Error-Rate by 33.4\% \cite{yang2020improvingdysarthricspeechintelligibility}. However, CycleGANs often produce artifacts especially in highly impaired speech and pixel level consistency can potentially be problematic and cause unrealistic images \cite{wang2024cycleganbettercycles}.  The DVC 3.1 system, which combines data augmentation with a StarGAN-VC backbone \cite{Zheng2023-ww} improved both ASR word recognition and listener ratings. Still, the quality of generated speech heavily depends on the synthetic data distribution and can degrade for highly variable input. More recently, diffusion-based voice conversion with Fuzzy Expectation Maximization (FEM) \cite{Hsu2024-cf} improved intelligibility and accuracy using soft clustering, but diffusion models are typically slow to sample. 

\subsubsection{Prior work: Speech to Text, Emotion, and Voice Cloning}
Prior work has explored speech to text pipelines on patients with dysarthria, marking the first step towards increased patient understanding \cite{shahamiri}. However, these methods have stopped at the transcription stage and have not progressed to sentiment classification or synthetic speech generation. In order to improve communication and understanding of dysarthric patients, it is crucial to provide this population with a way to communicate using acoustic features of their voice prior to their dysarthric diagnosis, enabling better understanding of sentiment and needs.

\section{Methods}
\subsubsection{Datasets}
We utilized four primary datasets in this study. The TORGO dataset \cite{torgo_dataset} provides paired audio samples and textual prompts from individuals with and without dysarthria, supporting analysis of articulatory impairments in English Speech. From this corpus, we accessed a subset of approximately 2,000 audio files (500 each for female non-dysarthric, female dysarthric, male dysarthric, and male non-dysarthric speakers) made available on Kaggle\cite{poojag718_dysarthria_dataset_2022}, and paired them with textual prompts sourced from a separate Kaggle dataset \cite{iamhungundji_dysarthria_detection_2023}. The UA Dysarthria dataset \cite{Vinotha_uaSpeechAll_2024} was used for severity classification and includes 11,436 speech spectrograms labeled across 4 severity levels: very low, low, medium, high. For Russian-language data, the Hyperkinetic Dysarthria Speech dataset \cite{mhantor_russian_voice_dataset_2023} was utilized, providing 2000 samples from both Dysarthric and non-dysarthric patients reciting the same phrase, along with corresponding prompts. Additionally, the Dysarthric German dataset \cite{czarnetzki_dysarthric_german_202} contributed 1,272 samples of Dysarthric German speech for crosslingual prediction.

\subsubsection{Initial Feature Extraction}

\begin{figure}[htbp]

\begin{minipage}[b]{1.0\linewidth}
  \centering
  \centerline{\includegraphics[width=8.5cm]{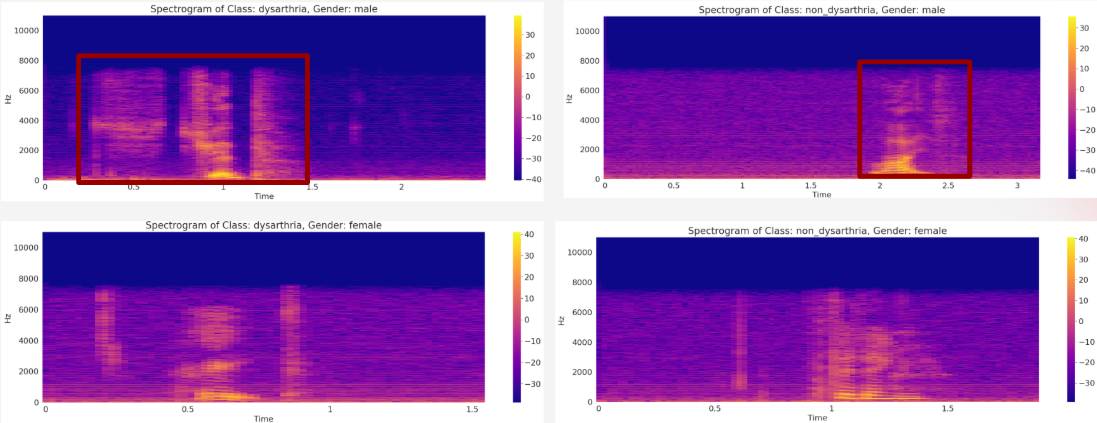}}
    \caption{Spectrogram visualizations of dysarthric and non-dysarthric speech across gender}
    \label{fig:spectrogramvis}
\end{minipage}
\end{figure}

To gain initial insight into speech patterns associated with dysarthria, we visualized spectrograms across gender and condition (dysarthric vs non-dysarthric). In the highlighted regions shown in Figure \ref{fig:spectrogramvis}, we observe that dysarthric speech tends to exhibit prolonged low-frequency spectral bands and reduced clarity, indicative of slurred articulation and irregular pacing. In contrast, non-dysarthric speech shows more distinct high-frequency bursts and cleaner articulation boundaries. 

\subsection{Dysarthria Detection Task}
Building on these qualitative differences, we extracted quantitative features for classification using Mel-Frequency Cepstral Coefficients (MFCCs), a widely used representation in speech processing. The pipeline begins with a Fast Fourier Transform to convert the raw audio into its frequency spectrum, followed by a logarithmic amplitude scaling that mimics human loudness perception. Mel scaling is then applied to emphasize perceptually relevant frequency bands. Finally, a Discrete Cosine Transform reduces dimensionality while preserving key spectral features. The resulting MFCCs capture articulatory and phonatory characteristics that are especially relevant for identifying dysarthric patterns.

\begin{figure}[htbp]

\begin{minipage}[b]{1.0\linewidth}
  \centering
  \centerline{\includegraphics[width=8.5cm]{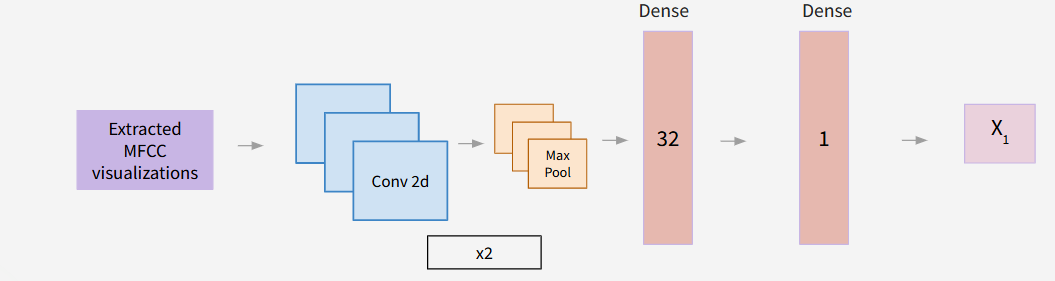}}
    \caption{Dysarthria Classification Model Architecture}
    \label{fig:modelarch1}
\end{minipage}
\end{figure}

To perform dysarthria detection from MFCC inputs, we adapted a simple 2D Convolutional Neural Network (CNN) architecture based on a publicly available Kaggle notebook. The model takes MFCC visualizations as input and passes them through two convolutional layers with max-pooling, followed by a dense layer with 32 units and a final output layer for binary classification. This architecture, as shown in Figure \ref{fig:modelarch1}, is effective at capturing time-frequency patterns relevant to dysarthric speech for our multilingual classification experiments. The model was trained for 50
epochs after which the training and validation loss converged.

\subsection{Severity Classification Task}
The second stage of our project involved classifying each dysarthric patient’s severity level as an indicator of how far along they are in their progression towards more serious diseases like ALS and Parkinson's. Accurate severity diagnosis can help a patient tailor their healthcare plan and treatment accordingly to better suit their needs \cite{improving_diagnosis_2015}. 

11436 spectrogram images of dysarthric audio were downloaded from Kaggle, resized to 128x128, converted to an array, and normalized. Class labels belonging to high severity (3036/11426), medium severity (2295/11426), low severity (2280/11426), and very low severity (3825/11426) were one-hot-encoded. The dataset split into train, test, and validation (70-20-10). 

\begin{figure}[htbp]

\begin{minipage}[b]{1.0\linewidth}
  \centering
  \centerline{\includegraphics[width=8.5cm]{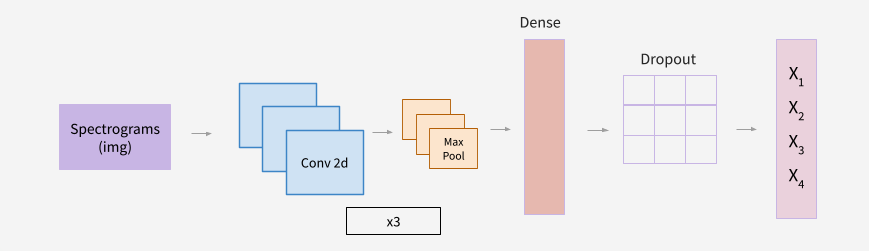}}
    \caption{ Severity Classification Model Architecture}
    \label{fig:modelarch2}
\end{minipage}
\end{figure}

A sequential model was defined with model architecture shown in Figure \ref{fig:modelarch2}. Three 2D convolutional layers were used to extract important features from the model, each of which were followed by a Max Pooling layer for dimensionality reduction. A dropout of 0.5 was specified to reduce overfitting and increase generalizability. The model was trained for 10 epochs after which the training and validation loss converged. 

\subsection{Multi-lingual clean speech synthesis}
The third component of the proposed framework is a two-stage pipeline to translate dysarthric speech into normal speech. Translating dysarthric speech into more understandable forms is crucial because reduced intelligibility severely limits a dysarthric patient’s ability to engage in daily conversations \cite{Page2022-rm}.

\subsubsection{Stage 1: Russian Dysarthric Speech to Normal Speech}

\begin{figure}[htbp]

\begin{minipage}[b]{1.0\linewidth}
  \centering
  \centerline{\includegraphics[width=8.5cm]{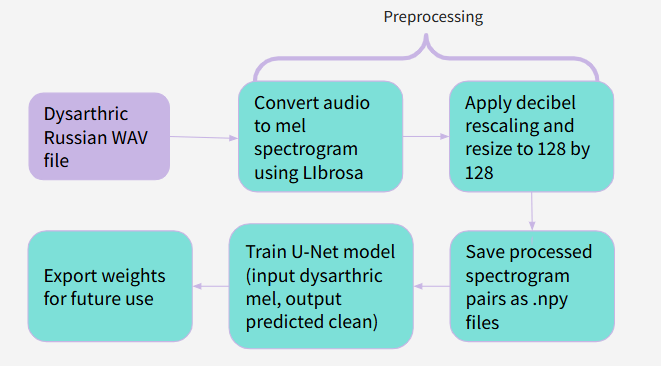}}
    \caption{Stage 1: Dysarthric Speech to Normal Speech (Russian)}
    \label{fig:phase1}
\end{minipage}
\end{figure}

In the first stage of our pipeline, we trained a U-Net model to map dysarthric speech to clean speech using Russian speech. As outlined in Figure \ref{fig:phase1}, raw .wav files were converted into mel spectrograms using Librosa, then rescaled in decibels and resized to a uniform 128x128 image for consistent model input. These paired spectrograms were saved as .npy files for training. The U-Net model was trained for 300 epochs, with a learning rate of 1e-4, to output a clean spectrogram from a dysarthric spectrogram, and the learned weights were exported. This step allows the model to learn structural transformations between distorted and healthy speech, that generalize across languages due to shared characteristics.  

\subsubsection{Stage 2: Adapting Russian Model to English Dysarthric Speech}

\begin{figure}[htbp]

\begin{minipage}[b]{1.0\linewidth}
  \centering
  \centerline{\includegraphics[width=8.5cm]{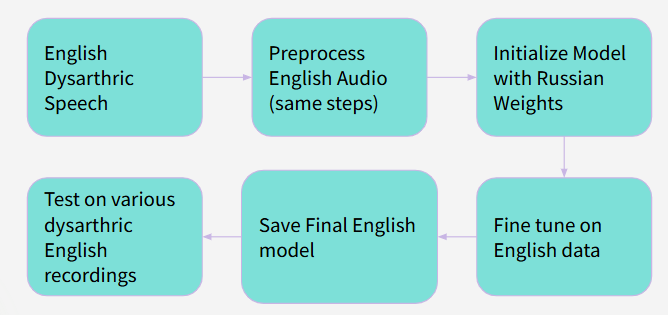}}
    \caption{Phase 2: Dysarthric Speech to Normal Speech (English)}
    \label{fig:phase2}
\end{minipage}
\end{figure}
Although the Torgo database contains substantial English dysarthric speech, we encountered a key challenge: very few clean–dysarthric pairs were spoken with the same text, which is necessary for paired spectrogram training. To construct a usable dataset, we manually filtered for matched male and female speakers saying the same sentences. After preprocessing, we identified only 190 matched female and 37 matched male samples. These were processed into mel spectrograms using the same pipeline as in Stage 1.
To address the limitations of training from scratch, we leveraged our U-Net model trained on the larger Russian dataset which had already learned to correct dysarthric distortions and followed the steps shown in Figure \ref{fig:phase2}.
We processed English dysarthric audio using the same preprocessing steps outlined in Phase 1, then initialized the model with Russian weights. We then fine tuned the model using the small available English dataset for 300 epochs and observed an improved performance over models trained from scratch. 
This cross-lingual transfer approach demonstrates how transfer learning can be used to compensate for scarcity of data and can be potentially applied to low-resource languages.

\subsection{Automatic Speech Recognition, Emotion Classification, and Voice Cloning} 

\begin{figure}[htbp]
\begin{minipage}[b]{1.0\linewidth}
  \centering
  \centerline{\includegraphics[width=8.5cm]{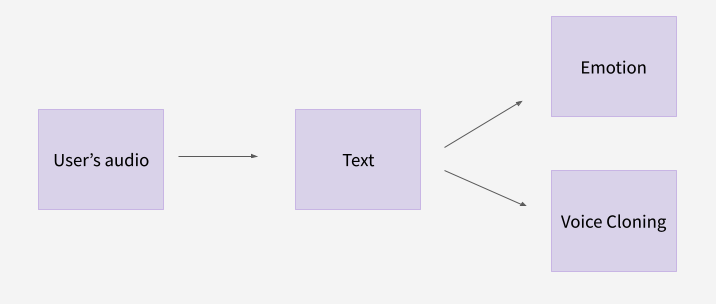}}
    \caption{Overall Speech to text, voice cloning, and emotion pipeline}
    \label{fig:overallpipeline}
\end{minipage}
\end{figure}

The overall pipeline for automatic speech recognition, emotion classification, and voice cloning is shown in Figure \ref{fig:overallpipeline}. We start by converting the audio to text, then use the text to classify emotion and perform voice cloning. 
\subsubsection{Speech to Text}

\begin{figure}[htbp]

\begin{minipage}[b]{1.0\linewidth}
  \centering
  \centerline{\includegraphics[width=8.5cm]{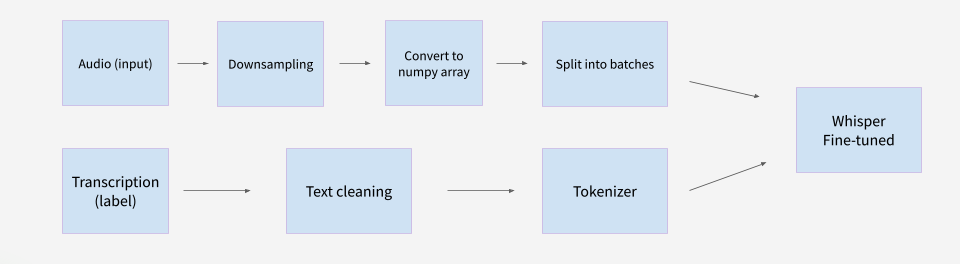}}
    \caption{Preprocessing pipeline for fine-tuning the Whisper model. The audio input undergoes downsampling, conversion to NumPy arrays, and batching. Corresponding transcriptions are cleaned and tokenized. Both processed inputs are then used to fine-tune the Whisper model for improved speech recognition performance.}
    \label{fig:s2t}
\end{minipage}
\end{figure}

The fourth component of our pipeline is a speech-to-text converter for dysarthric audio as shown in Figure \ref{fig:s2t}. This feature is crucial for increased communication for dysarthric patients, reducing the impact of dysarthria on their daily lives. 

The dataset was obtained from Kaggle, containing audio files of patients with and without dysarthria as well as their corresponding text. Our approach is to fine-tune an existing speech-to-text model on audio files of patients with dysarthria, thereby increasing the ability of these existing frameworks to comprehend slurred and slowed speech. 

After matching each audio file to their corresponding text, all instances of non-dysarthric patients were dropped to ensure that the model was only fine-tuned on patients with dysarthria, as the chosen models already performed well on clean speech. 

Three speech-to-text models were chosen for this application: Wave2Vec, Whisper, and Whisper Tiny. Transcriptions were cleaned to remove brackets and unnecessary spaces, and audio files were converted to numpy arrays and split into batches for quicker processing. The dataset was split with test size as 0.1, and transcriptions were converted using a tokenizer. Fine-tuning on Wave2Vec and Whisper proved to be difficult due to computational constraints and a large inference time, so Whisper Tiny was used for final mode fine-tuning. 

\subsubsection{Emotion Classification}

After obtaining speech to text results, an important component was adding an emotion classifier as shown in Figure \ref{fig:overallpipeline}, as sentiment is often lost in their reduced speech intelligibility and limitation in expressing nonverbal information \cite{alhinti2020recognising}. We used the pretrained Emotion English DistilRoBERTa-base transformer \cite{hartmann2022emotionenglish} to classify english text into 7 sentiments: anger, disgust, fear, joy, sadness, surprise, and neutral. Audio recordings were converted to text using the speech-to-text converter and then subsequently passed into the DistilRoBERTa-base model to infer the speaker’s emotional state.

\subsubsection{Voice Cloning}

\begin{figure}[htbp]
\begin{minipage}[b]{1.0\linewidth}
  \centering
  \centerline{\includegraphics[width=8.5cm]{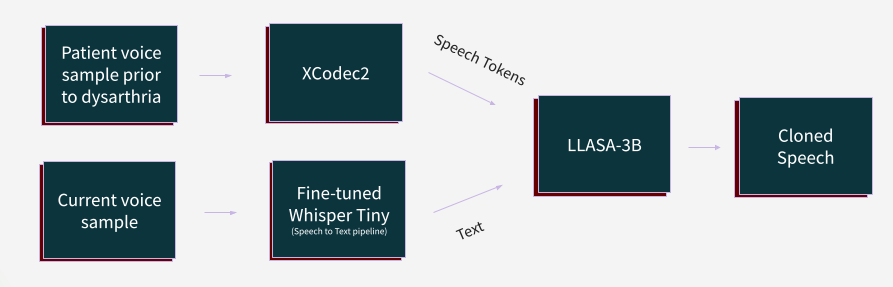}}
    \caption{Pipeline for voice cloning from dysarthric speech. The patient’s prior voice sample is encoded into speaker-specific tokens using XCodec2. Simultaneously, their current dysarthric speech is transcribed to text using a fine-tuned Whisper Tiny model. These speech tokens and text are jointly passed to the LLASA-3B model, which generates speech in the patient's original voice.}
    \label{fig:voiceclone}
\end{minipage}
\end{figure}

For patients with dysarthria, preserving their voice identity prior to their diseases is often crucial for better communication and understanding. Voice cloning is a process that reproduces a given dysarthric speech using input speech tokens from audio samples prior to the patient’s dysarthria as showcased in Figure \ref{fig:voiceclone}. 

Our voice-cloning text-to-speech (TTS) pipeline loads a user-provided sample audio (saying an arbitrary sentence) and resembles it to 16 hertz mono for input into an XCodec2 model. All reference code was taken from a pre-existing SOTA Text-to-speech and Zero Shot Voice cloning model \cite{billa2025llasa}. This speech codec model encodes a speaker’s vocal characteristics in a sequence of discrete tokens. The pipeline packages the speech tokens with the output from the speech-to-text pipeline, and feeds it into a LLASA-3B model which is fine-tuned to generate speech token sequences based on the text and speaker voice tokens. The generated speech tokens are then passed back onto the XCodec2 decoded to synthesize audio.

\section{Results}
\subsection{Dysarthria Detection}
\begin{figure}[htbp]
\centering
\includegraphics[width=0.45\textwidth]{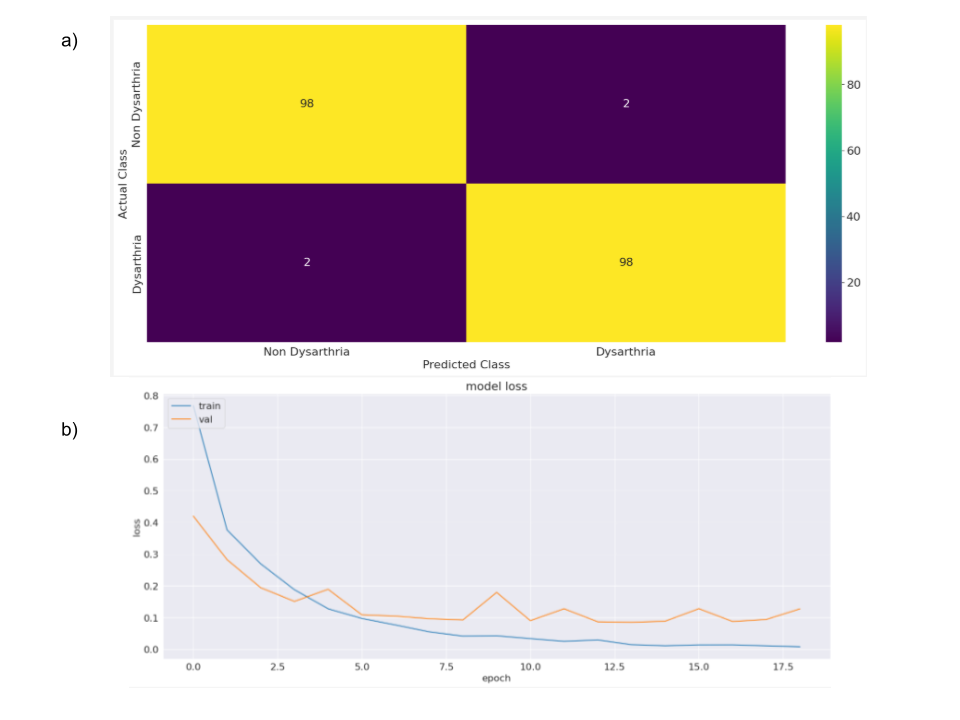}
\caption{(a) Training and validation loss curves for the model trained on the English (TORGO) dataset using MFCC features. The model shows stable convergence after ~8 epochs.
(b) Confusion matrix for the fine-tuned model on the Russian dataset, demonstrating high classification performance with 98\% accuracy in both dysarthric and non-dysarthric speech classes.}
\label{fig:combined}
\end{figure}

To detect the presence of dysarthria, we trained a binary classifier on MFCC features extracted from the TORGO English dataset. The model achieved a high accuracy of 97.5\%, and the training curve in Figure \ref{fig:combined}a shows stable convergence with minimal overfitting, supported by consistent validation loss. 

To evaluate cross-lingual generalization, we fine-tuned the English-trained model on German and Russian datasets. The model maintained high accuracy on both these languages as shown in Table \ref{tab:language_accuracy}. The confusion matrix in Figure \ref{fig:combined}b should that 98 out of 100 dysarthric and 98 out of 100 non-dysarthric samples were correctly identified in the Russian dataset, with only minimal misclassifications.

\begin{table}[htbp]
\centering
\begin{tabular}{|l|c|}
\hline
\textbf{Language} & \textbf{Accuracy (\%)} \\
\hline
English & 97.5 \\
German  & 96.8 \\
Russian & 99.7 \\
\hline
\end{tabular}
\vspace{0.5em} 
\caption{Accuracy of dysarthria detection across different languages.}
\label{tab:language_accuracy}
\end{table}

\subsection{Severity Classification}
\begin{figure}[htbp]
\centering
\includegraphics[width=0.45\textwidth]{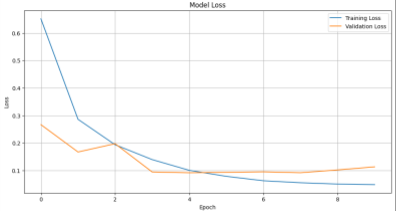}
\caption{Training and validation loss over epochs.}
\label{fig:loss1}
\end{figure}

\begin{figure}[htbp]
\centering
\includegraphics[width=0.45\textwidth]{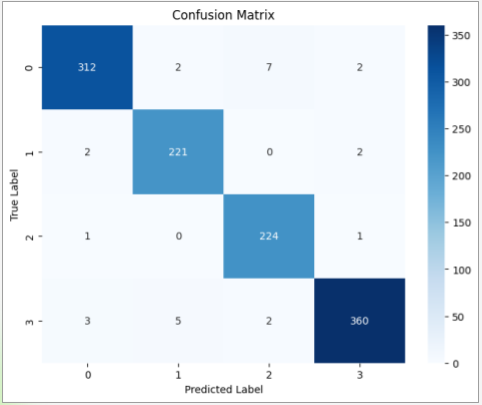}
\caption{Confusion matrix for severity classification.}
\label{fig:confusion}
\end{figure}

Our severity classifier was trained on spectrogram images of patients ranging across different severities of dysarthria. The model received a testing accuracy of 97.64\% as shown in Table~\ref{tab:accuracy}. The loss curves  in Figure~\ref{fig:loss1} show convergence after 8 epochs. The confusion matrix in Figure~\ref{fig:confusion} shows very few misclassified instances.

\begin{table}[htbp]
\centering
\begin{tabular}{|l|c|}
\hline
\textbf{Metric} & \textbf{Accuracy (\%)} \\
\hline
Training Accuracy   & 98.18 \\
Validation Accuracy & 97.38 \\
Testing Accuracy    & 97.64 \\
\hline
\end{tabular}
\vspace{0.5em} 
\caption{Model accuracy across training, validation, and testing datasets.}
\label{tab:accuracy}
\end{table}

\begin{figure}[htbp]
\centering
\includegraphics[width=0.45\textwidth]{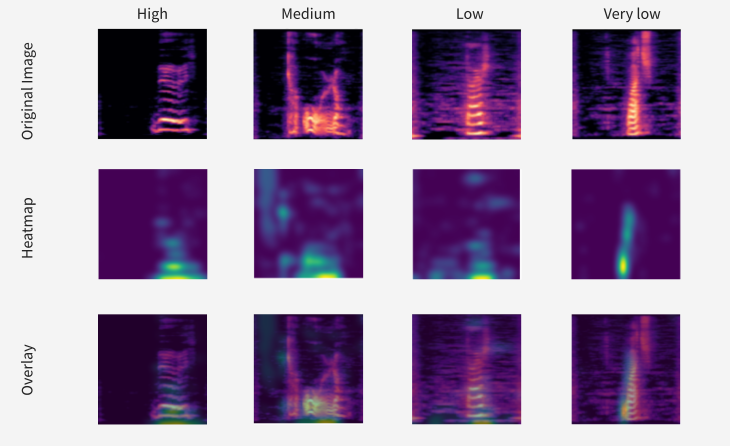}
\caption{Saliency Heatmap Showcasing Regions of Importance in Severity Classification}
\label{fig:gradcam}
\end{figure}

Another key feature of our model is interpretability. Grad-CAM heatmaps shown in Figure \ref{fig:gradcam} were produced by taking the gradients of the target class score with respect to the feature maps from a convolutional layer. Regions of importance as highlighted by the image are shown in yellow and green. 

\subsection{Speech to Speech Pipelines}

\subsubsection{Dysarthria Clean Speech Generation (Russian)}
\begin{figure}[htbp]
\centering
\includegraphics[width=0.45\textwidth]{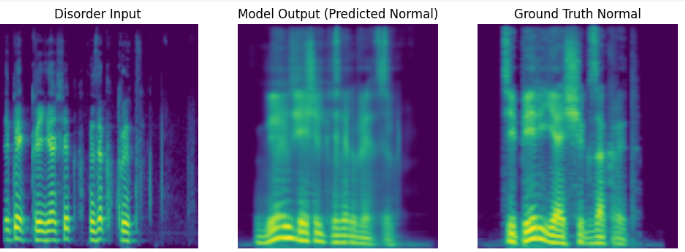}
\caption{Speech-to-speech transformation using Russian data. The left spectrogram shows dysarthric input speech, the middle displays the model’s predicted normal output, and the right shows the ground truth normal speech. The model output recovers key time-frequency structures associated with clarity and articulation.}
\label{fig:russian}
\end{figure}
Figure \ref{fig:russian} above displays the input dysarthric spectrograms, the U-Net model's predicted outputs, and the corresponding ground truth normal spectrograms. These visualizations confirm that the model successfully learns to denoise and restructure distorted speech patterns. While the outputs preserve the broad frequency distribution of the clean speech, they exhibit slight smoothing and blurring, likely due to the limited phase reconstruction during waveform conversion.

\subsubsection{Dysarthria Clean Speech Generation (Extended to English)}
\begin{figure}[htbp]
\centering
\includegraphics[width=0.45\textwidth]{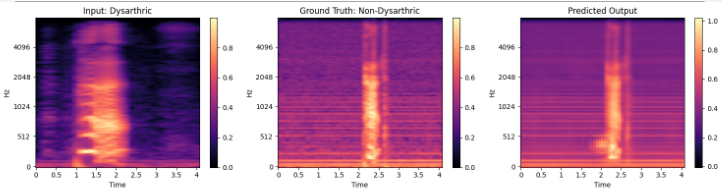}
\caption{Speech-to-speech transformation using English data. The left spectrogram shows input dysarthric speech, the center depicts the ground truth normal speech, and the right displays the model's predicted output. Despite limited training data, the model captures key spectral features, yielding a clearer and more intelligible output.}
\label{fig:english}
\end{figure}

Figure \ref{fig:english} shows the results of the  pretrained Russian U-Net model to generate normal speech from dysarthric speech fine tuned on the small English dataset. 

In Phase 2, the Russian-trained U-Net model was fine-tuned on a much smaller, carefully filtered English dataset. The figure showcases the input English dysarthric spectrograms, model predictions, and their matched ground truth normal counterparts.

\subsubsection{Comparison of Speech to Speech Results}
The model achieved a relatively low L1 training loss (0.03) and validation loss (0.06) shown in Table \ref{tab:unet_loss_comparison}. The fine-tuned model for the English dataset achieved a low train and test loss of 0.02 loss of 0.03 respectively.

\begin{table}[htbp]
\centering
\begin{tabular}{|l|c|c|}
\hline
\textbf{Model} & \textbf{U-Net (Russian)} & \makecell{\textbf{Pretrained Russian U-Net} \\ \textbf{finetuned on English}} \\
\hline
Training Loss & 0.03 & 0.02 \\
Test Loss     & 0.06 & 0.03 \\
\hline
\end{tabular}
\vspace{0.5em}
\caption{Comparison of training and test L1 losses across three model setups: (1) U-Net trained on Russian paired dysarthric–clean spectrograms, (2) Russian-trained U-Net fine-tuned on English data. Transfer learning via Russian pretraining leads to good performance on limited English data.}
\label{tab:unet_loss_comparison}
\end{table}

\subsection{Speech to Text}
\begin{figure}[htbp]
\centering
\includegraphics[width=0.45\textwidth]{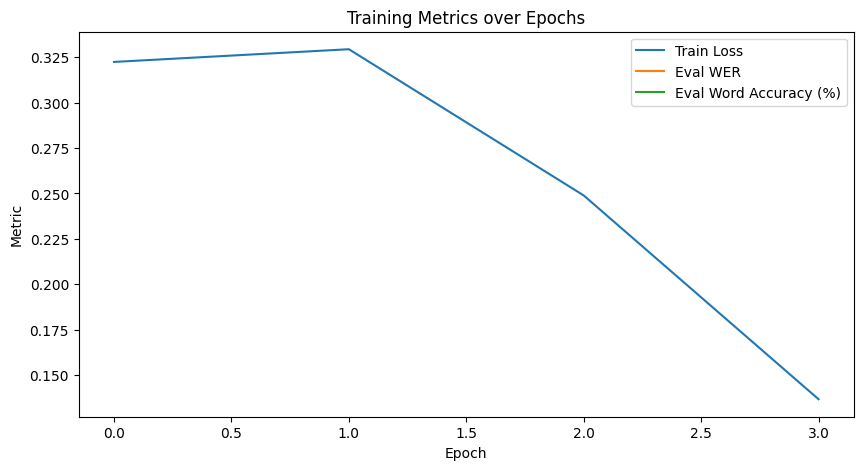}
\caption{Speech to Text Word Error Rate over Epochs}
\label{fig:loss}
\end{figure}

Our Speech-to-Text model received its best accuracy after 3 epochs. Table \ref{tab:wer_epochs} shows a training loss (word error rate) of 0.1367 towards the 3rd epoch, indicating an accuracy of 87.33\%. Figure \ref{fig:loss} indicates decreasing loss in only three epochs, a result of Whisper Tiny’s light framework designed for slurred speech. The output transcription can then be use for patient voice cloning, recreating the speaker’s original voice identity in what they say after being diagnosed with dysarthria.

\begin{table}[htbp]
\centering
\begin{tabular}{|c|c|}
\hline
\textbf{Epoch} & \textbf{Word Error Rate (WER)} \\
\hline
0 & 0.3224 \\
1 & 0.3294 \\
2 & 0.2489 \\
3 & 0.1367 \\
\hline
\end{tabular}
\vspace{0.5em} 
\caption{Word Error Rate (WER) across epochs during training.}
\label{tab:wer_epochs}
\end{table}

\subsection{Emotion}
Patients with dysarthria often have altered sentiment in their voice due to speech impairment. Model confidences shown in table \ref{tab:emotion_conf} indicate moderate levels of confidence across all emotions when analyzing speech transcriptions. Table \ref{tab:sentence_emotion} showcases sample sentences along with their corresponding emotion. 

\begin{table}[htbp]
\centering
\begin{tabular}{|c|c|}
\hline
\textbf{Emotion} & \textbf{Model Confidence} \\

\hline
Anger & 0.619131 \\
Disgust & 0.675264 \\
Fear & 0.625634 \\
Joy & 0.789678 \\
Neutral & 0.724674 \\
Sadness & 0.645108 \\
Surprise & 0.575870 \\
\hline
\end{tabular}
\vspace{0.5em} 
\caption{Confidence scores for Each Emotion Predicted by the DistilRoBERTa Model}
\label{tab:emotion_conf}
\end{table}

\begin{table}[htbp]
\centering
\begin{tabular}{|c|c|}
\hline
\textbf{Sentence} & \textbf{Predicted Emotion} \\

\hline
The snow blew into large drifts.
 & Anger \\
 Don't ask me to carry an oily rag like that.
 & Disgust \\
Before Thursday's exam, review every formula.
 & Fear \\
Bright sunshine shimmers on the ocean.
 & Joy \\
The store serves meals every day.
& Neutral \\
The family requests that flowers be omitted.
& Sadness \\
Yet he still thinks as swiftly as ever. & Surprise \\
\hline
\end{tabular}
\vspace{0.5em} 
\caption{Sample Sentences Corresponding to Each Emotion}
\label{tab:sentence_emotion}
\end{table}

\section{Discussion}
Overall, we were able to achieve high classification accuracy across English, Russian, and German demonstrating that our model is able to capture the acoustic patterns of Dysarthric speech while also generalizing to cross linguistic patterns. This cross-linguistic robustness is promising for improved dysarthria classification in other languages without as much available data.
Using a CNN for spectrogram-based severity detection also yielded promising and interpretable results, which enables early detection of dysarthria that a human examiner may not be able to pick up. Saliency heatmaps show that across all classes, the model is focusing on the lower harmonics, centered around the timepoints and frequencies that the signal lies in. This confirms the reason behind model prediction, as our model is able to look at a signal with interpretable results.

The results from our speech to speech model suggests that U-Net based spectrogram translation is promising for translating dysarthric speech to normal speech. By directly learning mappings from disordered to normalized speech spectrograms, the model is able to recover key time-frequency structures associated with clarity. While most existing systems rely on large matched datasets, our approach shows that using a pretrained architecture trained on Russian speech can be applied to low-resource languages. This highlights the potential of transfer learning for low resource languages, where collecting large paired datasets may not be feasible. 

ASR using transfer learning with Whisper Tiny effectively used the pretrained model and adjusted well to the limited dysarthria data using freezing and data augmentation. Whisper Tiny supports 99 languages and further research is needed for fine-tuning the transfer learning model to generalize to multiple languages, improving accessibility \cite{noauthor_undated-rw}. The main benefits of Whisper Tiny is that it is very robust, trained on 680,000 hours of multilingual data while also providing faster inferences than the two other transfer learning models tested (Wav2Vec and Whisper Tiny) \cite{Transcription_undated-ub}. 

Results demonstrate the effectiveness of our voice-cloning pipeline in reconstructing a patient’s original voice identity. While there is a significant degradation in voice identity between original voice and dysarthric voice, further improvements can  can be made by incorporating phonetic patterns such as how a speaker stresses syllables or transitions between vowels to improve voice intelligibility.

\begin{figure}[htbp]
\centering
\includegraphics[width=0.45\textwidth]{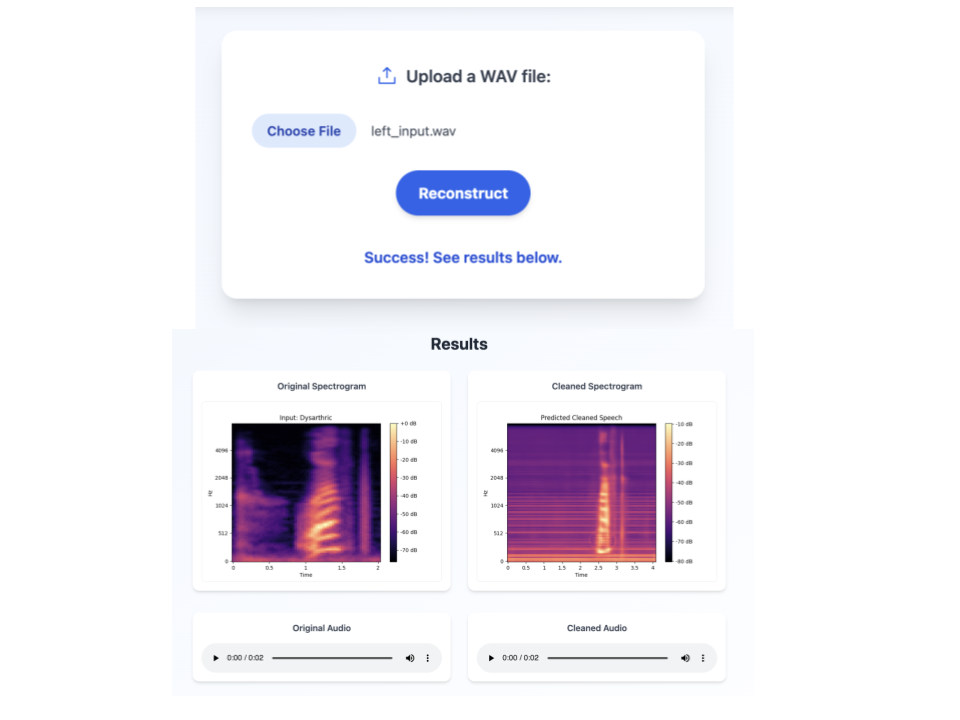}
\caption{Dysarthria Detection App built using HTML}
\label{fig:appinterface2}
\end{figure}
\subsection{Limitations}
While our framework demonstrates promising results across multiple tasks and languages, it has some limitations. First, our English speech-to-speech model was fine-tuned on a small paired dataset, which may restrict generalization across broader accents or sentence structures. Second, the emotion classifier was trained solely on clean transcriptions and may not fully capture emotion from more spontaneous or emotionally complex utterances. Finally, while our cross-lingual transfer approach worked well from Russian to English, its effectiveness across more structurally distant language pairs remains untested.
Future work includes expanding our dataset to include more diverse speakers and dialects, improving robustness to spontaneous speech, and testing the framework on additional low-resource languages. We also aim to refine the voice cloning pipeline to better preserve speaker identity over longer and more variable inputs.
\subsection{Mobile Launch}
To maximize accessibility we developed a web app using Flask and Javascript for backend and HTML and Tailwind CSS frontend. Images of the app interface are shown in Figure \ref{fig:appinterface2}. The Web app allows people to get timely, at-home dysarthria diagnosis results and easily use communication-aiding tools. In the future we hope to also develop a mobile app to improve accessibility.

\section{Conclusions and Future Work}
In this work, we present a multilingual framework for addressing the many dimensions of dysarthria, including detection, severity classification, speech-to-text transcription, clean speech generation, emotion classification, and voice cloning. Our models show high performance across English, Russian, and German datasets, demonstrating the potential for use in real-world multilingual settings.
To expand the reach of our framework, our next goal is to incorporate more low-resource languages where dysarthria diagnosis tools are especially scarce. We also aim to further reduce the Word Error Rate (WER) in our speech-to-text module by increasing dataset size, fine-tuning on more speech, and exploring multimodal data (e.g. combining acoustic features with visual inputs such as lip movements) to improve transcription accuracy. Our work is the foundation for a globally inclusive system for speech-based assistive technologies to bridge linguistic gaps and support communication and care for all patients with dysarthria.

\section*{Acknowledgment}
We are grateful to the Medlytics instructors and teaching assistants and MIT Beaver Works Summer Institute program directors for their guidance and encouragement throughout this project.

\bibliographystyle{IEEEbib}
\bibliography{references}

\end{document}